\documentclass[a4paper,11pt]{scrartcl}

\usepackage{amsmath,amssymb,amsthm}
\usepackage[british]{babel}
\usepackage{bbm}  
\usepackage{caption}
\usepackage[square,sort,comma,numbers]{natbib}
\usepackage{color}
\usepackage[top=2.3cm,bottom=2.3cm,left=2.5cm,right=2.5cm]{geometry}
\usepackage{graphicx}
\PassOptionsToPackage{hyphens}{url}\usepackage{hyperref}
\usepackage[utf8]{inputenc}
\usepackage{csquotes}  
\usepackage{nicefrac}
\usepackage[affil-it]{authblk}
\usepackage{microtype}      
\usepackage{subcaption}
\usepackage{tabularx}
\usepackage{float} 

\usepackage{amsmath,amsfonts,bm}

\def\1{\bm{1}}

\DeclareMathAlphabet{\mathsfit}{\encodingdefault}{\sfdefault}{m}{sl}
\SetMathAlphabet{\mathsfit}{bold}{\encodingdefault}{\sfdefault}{bx}{n}

\DeclareMathAlphabet{\mathpzc}{OT1}{pzc}{m}{it}

\definecolor{stefano}{RGB}{153, 0, 153}

\title{Epidemic mitigation by statistical inference from contact tracing data
}

\author[1]{Antoine Baker}
\author[2]{Indaco Biazzo}
\author[2,3,4]{Alfredo Braunstein}
\author[2]{Giovanni Catania}
\author[2,4]{Luca Dall'Asta}
\author[5]{Alessandro Ingrosso}
\author[1,7]{Florent Krzakala}
\author[2]{Fabio Mazza}
\author[1]{Marc M\'ezard}
\author[2]{Anna Paola Muntoni}
\author[1]{Maria Refinetti}
\author[6]{Stefano Sarao Mannelli}
\author[7]{Lenka Zdeborov\'a}

\affil[1]{Laboratoire de Physique de l’Ecole Normale Supérieure, Université PSL, CNRS, Sorbonne Université, Université Paris-Diderot, Sorbonne Paris Cité, Paris, France}
\affil[2]{Politecnico di Torino, Corso Duca degli Abruzzi 24, Torino, Italy}
\affil[3]{Italian Institute for Genomic Medicine, Via Nizza 52, Torino, Italy}
\affil[4]{Collegio Carlo Alberto, Via Real Collegio 30, Moncalieri, Italy \& INFN Sezione di Torino, Via P. Giuria 1, Torino, Italy}
\affil[5]{The Abdus Salam International Centre for Theoretical Physics, Strada Costiera 11, 34151 Trieste, Italy}
\affil[6]{Universit\'e Paris-Saclay, CNRS \& CEA, Institut de physique th\'eorique, 91191, Gif-sur-Yvette,France}
\affil[7]{\'Ecole Polytechnique F\'ed\'erale de Lausanne, 1015 Lausanne, Switzerland}

\begin{document}
\maketitle

Contact-tracing is an essential tool in order to mitigate the impact of pandemic such as the COVID-19.
In order to achieve efficient and scalable contact-tracing in real time, digital devices can play an important role. 
 While a lot of attention has been paid to analyzing the privacy and ethical risks of the associated mobile applications, so far much less research has been devoted to optimizing their performance and assessing their impact on the mitigation of the epidemic.

We develop Bayesian inference methods to estimate the risk that an individual is infected. This inference is based on the list of his recent contacts and their own risk levels, as well as personal information such as results of tests or presence of syndromes. We propose to use probabilistic risk estimation in order to optimize testing and quarantining strategies for the control of an epidemic. 
Our results show that in some range of epidemic spreading (typically when the manual tracing of all contacts of infected people becomes practically impossible, but before the fraction of infected people reaches the scale where a lock-down becomes unavoidable), this inference of individuals at risk could be an efficient way to mitigate the epidemic.
Our approaches translate into fully distributed algorithms that only require communication between individuals who have recently been in contact. Such communication may be encrypted and anonymized and thus compatible with privacy preserving standards.  We conclude that probabilistic risk estimation is capable to enhance performance of digital contact tracing and should be considered in the currently developed mobile applications. 

\newpage

\paragraph{Motivation}
\label{sec:Motivation}

One of the main tools public health authorities use to mitigate the spread of a pandemic, such as COVID-19, is the trace-test-isolate strategy. 
Identifying, calling, testing, and if needed quarantining the recent contacts of an individual who has just been tested positive is the standard route for limiting the transmission of a highly contagious virus. This standard strategy proves its efficacy at early stages of the epidemic, when the number of newly infected individuals is small enough to be manageable by reasonable-scale manual contact tracing infrastructures. However, it cannot be applied as such when the epidemic starts to spread faster, because the average number of contacts of a typical individual in the few days before he is tested positive can be large, not all contacts are with people known to the individual, and manual tracing incurs delays during which infected contacts keep on spreading the virus.

For these reasons, and taking into account the properties and parameters of the COVID-19 epidemic, digital contact tracing was convincingly argued to be a viable route to mitigation of COVID-19 and other similar epidemics \cite{Ferrettieabb6936}. Current mobile phone technology indeed enables automated, real-time proximity tracing between individuals and many works in this direction were initiated and deployed in past months \cite{bay2020bluetrace,AppleGoogle,troncoso2020decentralized,alsdurf2020covi,chan2020pact,chan2020pact}. With currently developed mobile applications, the distance and duration of a contact between two individuals can be estimated. Furthermore, contextual or health information about individuals can be included as well. 
This tracing can be used while preserving the privacy
of each individual's information, the level of privacy protection depending on the protocol. While many works have
been devoted, justifiably, to the compatibility of privacy and
tracing, see e.g. \cite{troncoso2020decentralized,bay2020bluetrace,chan2020pact,cho2020contact,raskar2020apps}, much less work is available concerning the assessment of the efficiency of tracing in mitigation of the pandemic. Most considered systems use the tracing data simply as a fast and scalable device to identify all recent contacts, in order to notify and perhaps isolate all of them.

In this paper we show that approximate Bayesian probabilistic inference
techniques allow to use the data exchanged by the tracing applications in order to provide highly accurate estimates of the probability that any given individual is infected. This estimate can then be used in order to focus the tests and other interventions on the group of individuals who have the largest probabilities of being infected, even if they do not show symptoms. The contact-tracing protocols that we propose require individuals that have been in contact in the recent past to be able to exchange messages about their risk level. When two individuals meet, they exchange a small amount of information (typically through Bluetooth). Later on, these individuals exchange messages carrying information about their current status, e.g. an increased risk due to presence of syndromes associated with the illness or due to their history of past contacts. Probabilistic inference then concatenates this information from all past contacts locally on the individuals phone and sends updates of the status to their contacts. 

We shall describe hereafter two concrete algorithms to perform the inference; one which is more accurate, based on Belief Propagation \cite{yedidia2003understanding}, and a second one which is a simpler approximation based on the so-called Mean-Field method. The latter requires smaller communication bandwidth between devices and could be potentially more privacy-friendly. Both the algorithms are inspired by our previous work \cite{Lokhov2014,lokhov2015dynamic,altarelli_bayesian_2014,altarelli_patient-zero_2014,altarelli_containing_2014,braunstein_inference_2016,bindi2017predicting,braunstein_network_2019}, and adapted to the present contact tracing problem. For a given testing capacity we show, through extensive simulation on realistic models for COVID-19 diffusion, that both methods allow to significantly reduce the impact of an outbreak and eventually contain the epidemics in many cases where standard tracing protocols fail to do so. We additionally evaluate robustness of the methods to presence of false negative tests as well as partial adoption of the contact-tracing mobile applications.

\paragraph{Related work}
While most considered systems use the tracing data simply as a fast and scalable device to identify all recent contacts, works aiming at estimating the risk of infection appeared recently. These include a machine learning based risk estimation proposed in \cite{alsdurf2020covi}, that provides only limited validation of the approach even on data specific to the privacy preserving protocol, and is thus difficult to directly compare to our approach. Preliminary version of our work was first presented in Ellis COVID-19 workshop \cite{Ellis_conf}. An approach similar to our MF algorithms was proposed by \cite{ViraTrace} using Monte Carlo sampling to estimate the corresponding probabilities. This work does not provide validation of the approach involving the control of the epidemic. Another recent work estimating the risks from the tracing data is \cite{herbrich2020crisp} that is in the spirit similar to our work, it uses Monte Carlo sampling based estimations of the risks. The authors of \cite{herbrich2020crisp} evaluate their approach only on data that come from the model that is assumed in the inference algorithm. A key aspect of our work is that we test on data coming from a much more complex model than assumed when designing the inference algorithm \cite{hinch_2020}. We believe that this is crucially important for eventual validation on real world contact data, which are not available to us at this point. We note that the authors of \cite{herbrich2020crisp} evaluate their approach on networks up to 10k individuals, compared to 500k used in our simulations. As is common with Monte-Carlo schemes, convergence properties could significantly deteriorate with system size. The  lack of separation between the epidemic generating model and the inference procedure in the implementation of \cite{herbrich2020crisp} makes it difficult for us to compare directly to our approach, and we thus leave it for future work. The python code used for our simulations \cite{epidemicmitigation} is modular and comparisons with other inference procedures can be performed by adding new modules.

%

\paragraph{Scheme of propagation}
\label{sec:Propagation}

A convincing validation of any individual-level intervention policy requires extensive simulations by means of sufficiently detailed agent-based models. In such models, at each time, a given individual is in a state that belongs to a finite set of possible states, like for instance susceptible, exposed, infected-asymptomatic, infected-asymptomatic, in ICU, recovered, or dead.    
The most accurate mathematical descriptions of COVID-19 epidemic propagation are based on  complex multi-state compartment models, in which infected individuals are not immediately contagious upon infection, may be asymptomatic or develop mild/severe symptoms with some delay, and the ages, households and workplaces are also taken into account. Even though the long-term effects of SARS-CoV-2 infection are still under study, it seems reasonable to assume that some level of immunity is developed with recovery, so that the individual progression through the epidemic compartments is not recurrent (a recovered person typically does not become infected again). The observation of non-trivial distributions of incubation and recovery times as well as that of time-dependent viral transmission capacity \cite{2020BentoutParameterEstimation,franco2020covid19,fintzi2020using,KEFAYATI2020,KEFAYATI2020} indicate that the most realistic models for SARS-CoV-2 infection clearly depart from the simplest, and largely adopted, Markovian epidemic models. 
In addition, such models also provide representations of the time-varying contact network over which viral transmissions occur, some including real-world mobility data \cite{lorch_spatiotemporal_2020} or computer-generated synthetic surrogates \cite{ferretti_quantifying_2020,hinch_2020}. In particular, the model in Ref.~\cite{hinch_2020} simulates the spread of COVID-19 in urban age-stratified populations with a multi-layer contact network (see also Supplemental Material \ref{app:OpenABM} for details). 

\paragraph{Bayesian Epidemic Tracing}

Information regarding the status of tested or symptomatic individuals can be used in different ways within a contact tracing procedure. 
In the simplest situation, 
the observation of an infected individual involves tracing (a fraction of) his recent contacts in order to prevent/contain further transmissions of the infection. It is also possible 
to infer transmission chains and detect the parent cases which are the origin of the infection detected in an individual. 
In a Bayesian approach, this is made possible assuming as a prior a particular probabilistic model of epidemic propagation and using it to define a likelihood function conditioned on the evidences coming from observational data, i.e. tests (PCR and/or serology) and self-reported symptoms from a fraction of the individuals. As we shall show, the adopted prior inferential model  provides the mathematical framework for developing risk assessment, but it does not need to reflect the real epidemic spread in all its details in order to allow for valuable inference. Indeed, for epidemic propagation generated with complex agent-based models along the lines described above, we show that the approximate computation of local probability marginals can be effectively obtained using  as a prior much simpler inferential models. In practice in this paper we use a simple agent-based Susceptible-Infected-Recovered (SIR) model. 
We emphasize that, depending on the approximate inference technique used for computing such probability marginals, some of the  ingredients of realistic epidemic propagation can be reintroduced, such as non-Markovian evolution between states, time-dependent infectiousness or more compartments.

Here we propose two distributed algorithms for risk estimation from contact tracing data, which are both derived within a Bayesian framework and based on a message-passing principle: Mean-Field (MF) algorithm and  Belief Propagation (BP) algorithm. They both provide estimates of the time-dependent local probability marginals of being infected which can be used to estimate the risk level of each individual; this risk of being infected can in turn be used to implement a sanitary protocol, like suggesting higher risk individuals  to be tested and/or quarantined. The main differences between these methods, which are presented in the next section, are in the accuracy of the approximation of the epidemic dynamics 
and in the way in which observational data are incorporated into the probabilistic model.

\paragraph{Methods}\label{sec:Methods}
We use the Bayesian approach that is based on a prior description of the epidemic process, and a method to include the information from observations.
We present here, and use in this paper, a prior description based on a discrete-time Markov chain corresponding to the SIR model. It can be generalized to non-Markovian, continuous-time, and other model settings.

Let $x_i^t$ be the state of an individual at time $t$ (it is convenient to think of $t$ as a number of days), with $x_{i}^{t}\in X$ and $X$ a finite set of epidemic states. We use $X=\{S,I,R\}$ for the case in which the susceptible ($S$), infected ($I$) and recovered ($R$) individual states are considered. The state of individual $i$ at time $t$, $x_{i}^{t}$, depends on her state at the previous time, and on the states at time $t-1$ of all the individuals $j$ that she has met between times $t-1$ and $t$. We denote by $\partial i (t)$ this set of individuals, and by $x_{\partial i}
^{t-1}=\{x_j^{t-1}\}$ for $j \in \partial i (t)$. Then  
$p\left(x_{i}^{t}|x_{\partial i}^{t-1},x_{i}^{t-1}\right)$ is the probability of individual transitions for $i$ occurring between time $t-1$ and time $t$.
For the SIR model, this probability depends on the following parameters: 
\begin{itemize}
    \item 
the recovery rate $\mu_i$, defining the daily probability that the infected individual $i$ moves to the recovered state 'R';
\item 
the transmission rates $\{\lambda_{k\to i}(t)\}_{k \in \partial i(t)}$, which are the probability of infection from an infected $k$ to a susceptible $i$ on day $t$.
\end{itemize}

Let $\mathbf{x}=\left\{ x_{i}^{t}\right\} _{i=1,\dots,N}^{t=0,\dots,T}$
be a collective time-trajectory generated by the epidemic.  
The prior probability associated with this trajectory is defined by 
\begin{align}
p\left(\mathbf{x}\right) & =\prod_{i}p\left(x_{i}^{0}\right)\prod_{t=1}^{T}p\left(x_{i}^{t}|x_{\partial i}^{t-1},x_{i}^{t-1}\right)\label{eq:markov} \, ,
\end{align}
 where we assumed a factorized probability of initial state  $\mathbf{x}^{0}=\{x_i^0\}_{i=1,\dots,N}$, i.e. $p\left(\mathbf{x}^{0}\right)=\prod_{i}p\left(x_{i}^{0}\right)$.

We can now include the effects of observations. Given a set $O$ of  observations $\mathcal{O}=\left\{ \mathcal{O}_{r}\right\}_{r\in O}$, where each observation $r$ provides some information on the state of an individual at a given time (as the result of tests, or of individual symptoms), and assuming that these observations are statistically independent,
the posterior probability of the trajectory $\mathbf{x}$ can be expressed using Bayes theorem as 
\begin{align}\label{eq:prob_model2}
p\left(\mathbf{x}|\mathcal{O}\right)  &=\frac{1}{p\left(\mathcal{O}\right)}p\left(\mathbf{x}\right)\prod_{r}p\left(\mathcal{O}_{r}|\mathbf{x}\right) \nonumber \\
&=\frac{1}{p\left(\mathcal{O}\right)}\prod_{i}p\left(x_{i}^{0}\right)\prod_{t=1}^{T}p\left(x_{i}^{t}|x_{\partial i}^{t-1},x_{i}^{t-1}\right)\prod_{r}p\left(\mathcal{O}_{r}|\mathbf{x}\right).
\end{align}

For the BP approach, we start from the posterior at \eqref{eq:prob_model2} and remark that it can be written as:
\begin{align}
p\left(\mathbf{x}|\mathcal{O}\right) 
& =\frac{1}{Z}\prod_{i}\psi_{i}\left(\mathbf{x}_{i},\mathbf{x}_{\partial i}\right). \label{eq:graphical_model}
\end{align}
Belief Propagation \cite{yedidia2003understanding} can then be used to estimate marginal posterior probabilities from \eqref{eq:graphical_model}. However, a straightforward factor graph representation \cite{mezard2009information,pearl1982reverend} of \eqref{eq:graphical_model} with $\{\mathbf{x}_i\}$ as variable nodes and the compatibility functions $\{\psi_{i}\left(\mathbf{x}_{i},\mathbf{x}_{\partial i}\right)\}$ as factor nodes, contains many short loops, so that the corresponding BP equations would not be exact even when the underlying contact network is acyclic. We instead construct a factor graph representation that closely reflects the topological structure of the contact network by associating the individual trajectories of a pair of individuals in contact, and involves BP messages $m_{ij}\left(\mathbf{x}_{i},\mathbf{x}_{j}\right)$ for pairs of trajectories. The corresponding BP fixed-point system for $\{m_{ij}\left(\mathbf{x}_{i},\mathbf{x}_{j}\right)\}_{(ij)\in E}$ is solved by iteration.  
This formalism has been employed for large-deviation analyses of a class of dynamical processes including applications to epidemics \cite{altarelli_large_2013,altarelli_optimizing_2013,altarelli_bayesian_2014,altarelli_patient-zero_2014,braunstein_inference_2016}, in particular regarding the patient zero problem and the inference of causality chains of infection. We extended here previous works to deal with non-Markovian processes and to make it computationally efficient through a limited time-window approximation (see Supplemental Material).

Restricted to Markovian epidemic models, a simpler, Mean-Field approximation can be devised starting from \eqref{eq:markov}. It is based on assuming that $p\left(\mathbf{x}^{t}\right)\approx\prod_{i}p\left(x_{i}^{t}\right)$, so that 
\begin{equation}
p\left(x_i^{t+1}\right)\approx\sum_{x^t_i,\mathbf{x}^t_{\partial i}}p\left(x_{i}^{t+1}|\mathbf{x}_{\partial i}^{t},x_{i}^{t}\right)p\left(x_{i}^{t}\right)\prod_{j\in\partial i}p(x_j^t).
\end{equation}
Thanks to this factorization, one can write closed  equations for the evolution of the individual probabilities $p(x_i^t)$ for the simple prior model (\ref{eq:markov}) along the same spirit as presented in \cite{Lokhov2014,lokhov2015dynamic}, (see Supplemental Material).

For risk inference, we need to estimate the $p(x_i^t|O)$ in the full model (\ref{eq:prob_model2}) that includes the observations. We show in the Supplemental Material a heuristic that incorporates the presence of  observations done at a time $t_{obs}$ into the Mean-Field equations. The algorithm propagates the information on the population from $t_{MF}$ days before the current time and it simply takes into account the following facts:
\begin{itemize}
\item
If an individual is tested 'S' at time $t_{\text{obs}}$, it has been 'S' at all previous times
\item
If an individual is tested 'R' at time $t_{\text{obs}}$, it will be 'R' at all following times
\item
If an individual is tested 'I' at time $t_{\text{obs}}$, we assume that he has been 'I' at times  $[t_{\text{obs}}-\tau,t_{\text {obs}}]$, where $\tau$, the typical time between infection and observation, is a parameter of the algorithm.
\end{itemize}

%
%
%

%
%
%

\paragraph{Results}
\label{sec:results}
We test the inference of risk on two types of epidemic spreading and contact networks. 
\begin{itemize}
    \item SIR spreading model on proximity-based random network: This is a simple SIR-model-based propagation in a population of $N$ individuals, where the graph of contact is updated dynamically at each step as follows. The individuals are  distributed uniformly in a square of side $\sqrt{N}$, and at each time step a contact can be established between two individuals $i$ and $j$ with a probability $e^{-d_{ij}/\ell}$, where $d_{ij}$ is the Euclidean distance between the points and $\ell$ is a parameter that controls the density of the contact graph. We shall call this the geometric contact model.
\item Oxford OpenABM model: The second model is a much more realistic epidemic spread model \cite{Ferrettieabb6936}, which is aimed at capturing essential features of the contacts in real populations as well as the real epidemiology of COVID-19, we call it the OpenABM model. In the absence of sufficiently detailed real world data, we view the data from the OpenABM model as realistic and our main point is to demonstrate that even though the proposed inference procedures do not capture most of the details and complexity of this model, they still work and provide large improvement over competing current contact tracing methods.
\end{itemize}

For the OpenABM model, we use in MF inference an extremely simplified hypothesis of equal recovery rates, $\mu_i=\mu$, and transmission probabilities only divided into two classes (inter-household contacts, with $\lambda_{k\to i}(t)=\lambda$ and intra-household contacts, with $\lambda_{k\to i}(t)=2\lambda$). The values of  parameters $\mu$ and $\lambda$ are chosen on the basis of population averages; they could also be inferred from data. Note that arbitrarily heterogeneous parameters can be used if more information is available, such as the duration of a contact. It is important to notice that the MF algorithms that we derive from this simple model turn out to be very efficient at predicting the risk of an individual to be infected, even in sophisticated propagation models that involve individual and time-dependent rates $\mu_i$ and $ \lambda_{k\to i}(t)$. The BP inference model is slightly more complex (see Section \ref{subsec:BPparams} for details), although still much simpler than the original OpenABM propagation model.

In both models, we start the simulation at time $0$ with everybody in the susceptible state $S$ except a small number of infected individuals. The number of these "patients-zero" will be specified in the following for each case.

We apply the following testing protocol. We observe a fraction of symptomatic individuals at the day of symptoms. After a fixed number of days ($t_{\text start}$) 
the interventions start. Every day, we perform 
 a fixed amount $n_r$ of tests to the top individuals ranked as having the largest probabilities of being infected, according to the different risk estimation strategies. We assume that the result of the test is available on the same time step (day) and is included in the observations used to adjust the probabilities of risk on the next time step (day). 

Besides BP and MF risk estimation, the ranking strategies considered for comparison are:
\begin{itemize}
    \item Random Guessing (RG): The $n_r$ individuals on which the tests are performed are randomly chosen among the individuals that were not previously tested positive.
    \item Contact Tracing (CT): One ranks the individuals who have not been tested positive previously
    according to the number of contacts with confirmed positive individuals during the time interval $[t-\tau,t[$, and tests the $n_r$ individuals with the largest number of contacts. This is what would be possible to implement with the currently deployed mobile applications. 
\end{itemize}
For BP and MF inference, the ranking is done as follows:
\begin{itemize}
    \item Belief Propagation (BP) and Mean-Field (MF): One uses the algorithm (BP or MF) in order to estimate the probabilities $q_i^t=p(x_i^t=I)$ of being infected at time $t$. Individuals who have not been tested positive previously are ranked according to their risk $q_i^t$, the $n_r$ individuals with largest risk are tested. For BP, the rank is computed as the probability of infection in the last $\delta_{\text{rank}}$ days. Prioritizing recent infections can be more effective as it helps containing the "boundary" of an ongoing outbreak. 
\end{itemize}

\begin{figure}
\centering
		\includegraphics[width=0.5\linewidth]{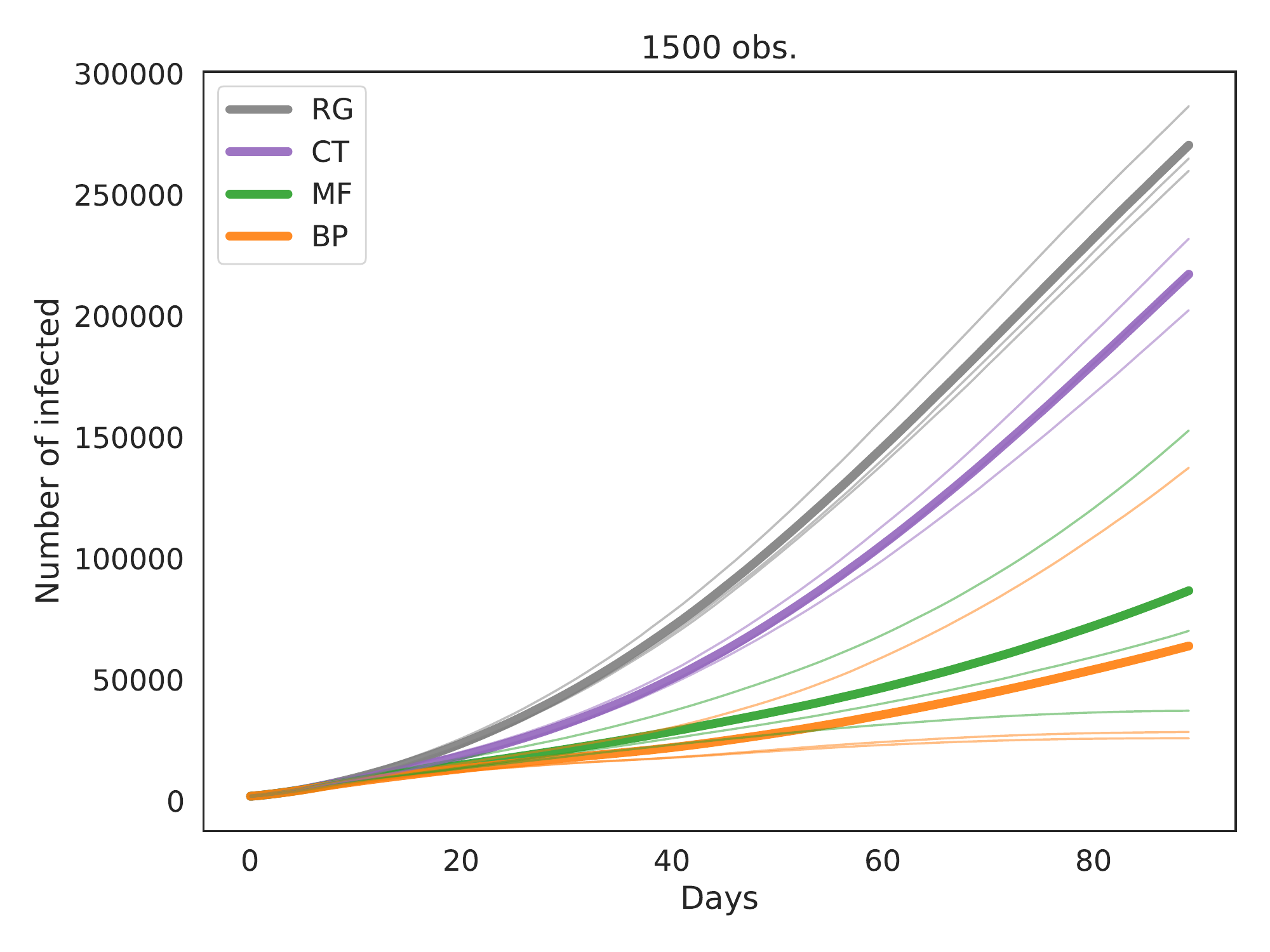}
	\caption{Spreading of the epidemics in a 2D geometric graph with 6 contacts on average per day (scale of the graph is 1) and $500\,000$ individuals. The parameters of the forward simulation of the SIR model are the same as used by the inference algorithms: $\lambda=0.05, \mu=0.02$. In the plot we show the average numbers (bold lines) of infected individuals versus time of simulations among three different realizations (thin lines) with 200 patients zero. The system freely evolves for the first 10 days, then interventions start. We observe $50\%$ of the infected individuals 5 days after their infection. We perform 1500 tests every day according to the ranking given by the algorithms. The observed infected individuals are quarantined
	. The MF parameters are $\tau = 5$, $t_{\text{MF}}=15$.
	}
	\label{fig:geometric_graph2}
\end{figure}

 We compare test-guided containment strategies based on MF, BP, RG and CT in a scenario where quarantines are put in place when tested individuals result infected. We show that
 BP and MF-based methods are able to predict infections and control the epidemic considerably more successfully than the classic contact tracing strategy. Implementation of the MF and BP risk estimation algorithms and all the tests that follow can be found at \cite{epidemicmitigation}.

 We evaluate the proposed framework in a pessimistic regime with 200 simultaneous independent outbreaks that are discovered after ten days.  
 We start with a simulation of the proximity-based random network. 
 Figure \ref{fig:geometric_graph2} shows the development of the epidemic over three months in a population of $500\,000$ individuals, starting from 200 patients-zero, and performing $n_r=1500$ tests per day. In spite of rather large fluctuations from run to run, one sees a very clear signal indicating that the proposed inference methods, MF and BP, largely improve upon the usual CT, which is itself better than RG. The best inference method is clearly BP, but the simpler MF, which is less demanding in terms of the amount of information exchanged between individuals, and therefore easier to protect for better privacy, is also quite successful. Even in this pessimistic regime both risk inference methods allow to slow down the epidemic spread by more than a month compared to classic contact tracing.

\begin{figure}
    \centering
    \includegraphics[width=1.0\linewidth]{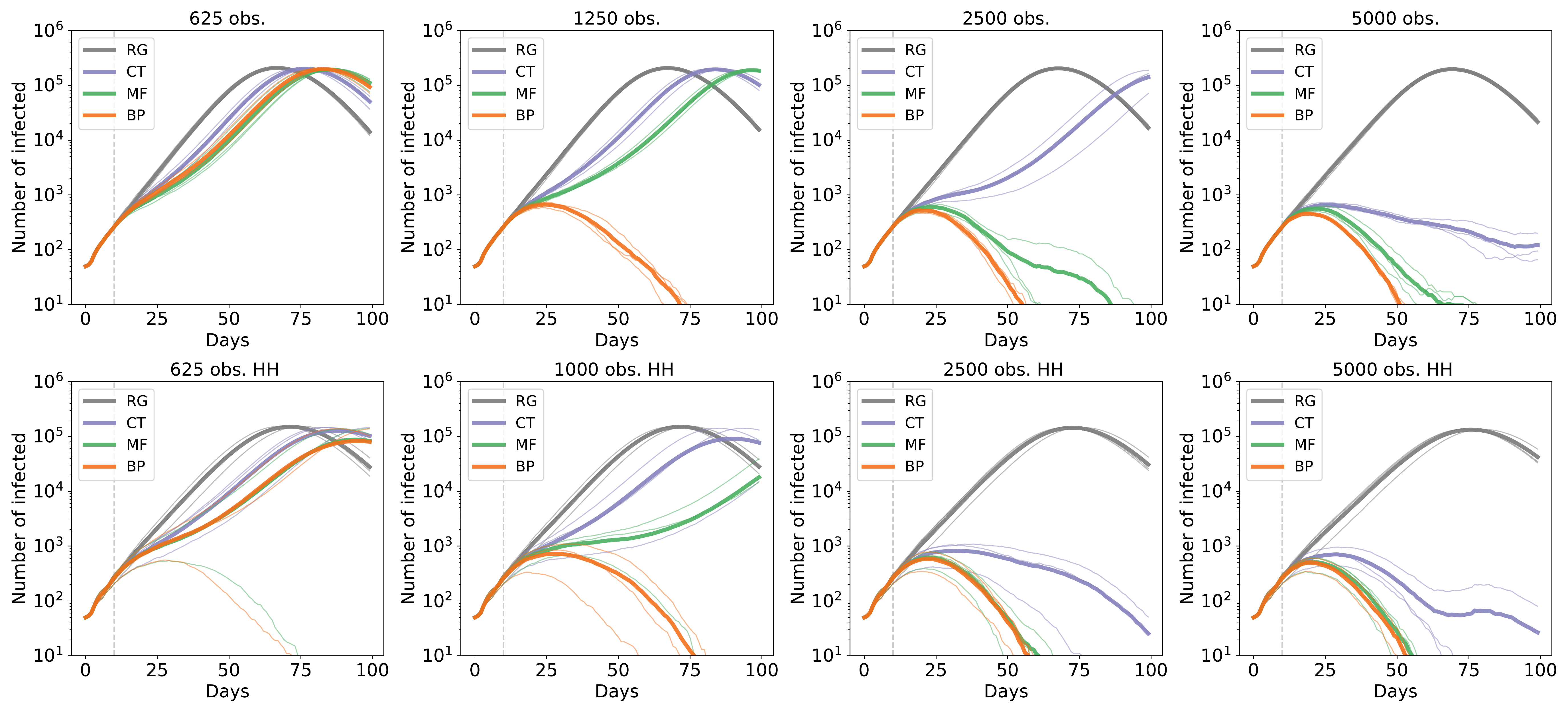}
    \caption{Effect of the control strategy on the epidemic spreading, according to the OpenABM model, in a population of $500\,000$ individuals. In all the panels we show the number of infected individuals in a time window of 100 days when interventions are applied starting from day 10. The number of patients zero here is set to 50. Thin lines represent the results for single instances of the epidemics, while the thick line is the average among the different realizations. We compare the effect of an increasing number of available medical tests per day (from left to right), performed to the individuals at highest risk as evaluated by the corresponding strategy (RG, CT, MF and BP). The top panels depict a scenario where only tested positive individuals are confined, limiting their contacts to the cohabitants, while the bottom panels show how the number of infected individuals change if the entire household is quarantined whenever an infected member is detected. 
    \label{fig:res_obs_1a}}
\end{figure}
 
 In this first test we assumed that the model used for BP and MF inference coincides exactly with underlying epidemic propagation model (SIR), which is overly optimistic. A much more stringent test has been performed on the more realistic OpenABM COVID-19 model. 
 In this framework, each infected individual can either be asymptomatic or show symptoms of various degree (mild or severe). As in a realistic setting, we will consider an additional source of information: individuals that show severe symptoms are immediately quarantined when symptoms emerge (typically 5 days after infection) or hospitalized. In addition, half of the mildly symptomatic individuals is assumed to self-report and self-isolate as well. No direct information is available on asymptomatic (or pre-symptomatic) infected individuals, their detection is possible only through contact tracing. 

We mimic a post lock-down scenario where only a small fraction of individuals is initially infected, i.e. few tens of patients zero in a population of 500 thousand individuals that all employ a contact-tracing application. The epidemic dynamics freely evolves according to the OpenABM model \cite{hinch_2020} for ten days and then a number of individuals with the highest infection risk, assessed by RG, CT, MF, or BP, is tested on a daily basis. The MF algorithm assumes a Markovian SIR spreading with parameters $\lambda=0.02$ and $\mu=1/12$, and has parameters $\tau=5$ and $t_{\text{MF}}=10$. See Section \ref{subsec:BPparams} for details on the BP parameters. In these simulations, the result of the medical tests is assumed to be exact; errors in tests will be addressed below. 
The number of medical tests associated with the individuals detected by the mobile application is fixed while there is no limitation on those performed to the fraction of symptomatic people presented above. The original contact dynamics is then modified in agreement to two alternative strategies: tested positive are confined and can have contacts only with their cohabitants or, whenever one person results positive to the medical test, all the households are quarantined without being tested. 

Figure \ref{fig:res_obs_1a} shows the number of infected individuals in a time interval of 100 days when the number of initial infections is 50 and the intervention starts after 10 days. 
In the top panels, we show the results for three independent realizations of the epidemics in the case where tested positive individuals only are quarantined, while in the bottom panels we show the results for a more restrictive intervention scenario in which all the households are confined. The number of available tests per day increases from 625 to 5000 (from left to the right panels). The lines are colored according to the adopted ranking strategy and the thick lines show the mean number of infected individuals mediated on the three instances. The results suggest that for both the inference strategies the size of the epidemics is significantly reduced if compared to the random testing and also to the classic contact tracing, even when few tests are available. We remark the behavior of the BP-based strategy when 1000 (1250 when the confinement is not extended to the households) daily medical tests are performed: the confinement of the people inferred by this method suffices to stop the epidemic after 75 days. The MF-based strategy performs notably better than CT and it achieves similar performances to BP when the number of daily observations is large.

In Figure \ref{fig:res_obs_1b} of the Supplemental Material we show the number of infected individuals for a time window of 100 days in, to some extent, a different scenario, that is when the containment measures are applied earlier in time, after a week from the beginning of the epidemics, and the size of the epidemics at initial time is smaller than that examined in Figure \ref{fig:res_obs_1a}. 
 
 \paragraph{ Robustness of the inference}
 
 In the previous section we investigated how several intervention protocols (differing in the treatment of the households and the number of available tests) control realistic epidemics when paired to the considered risk assessment strategies (RG, CT, MF and BP). However, some of the conditions assumed in that section 
 are not realistic. In reality, the sensitivity of medical tests is not 100\% and it is to be expected that only a fraction of the population will adopt the app, so that not all contacts are detectable. In this section, we shall address these two issues, focusing on the more realistic OpenABM model.

\begin{figure}[h]
    \centering
    \includegraphics[width=1.0\linewidth]{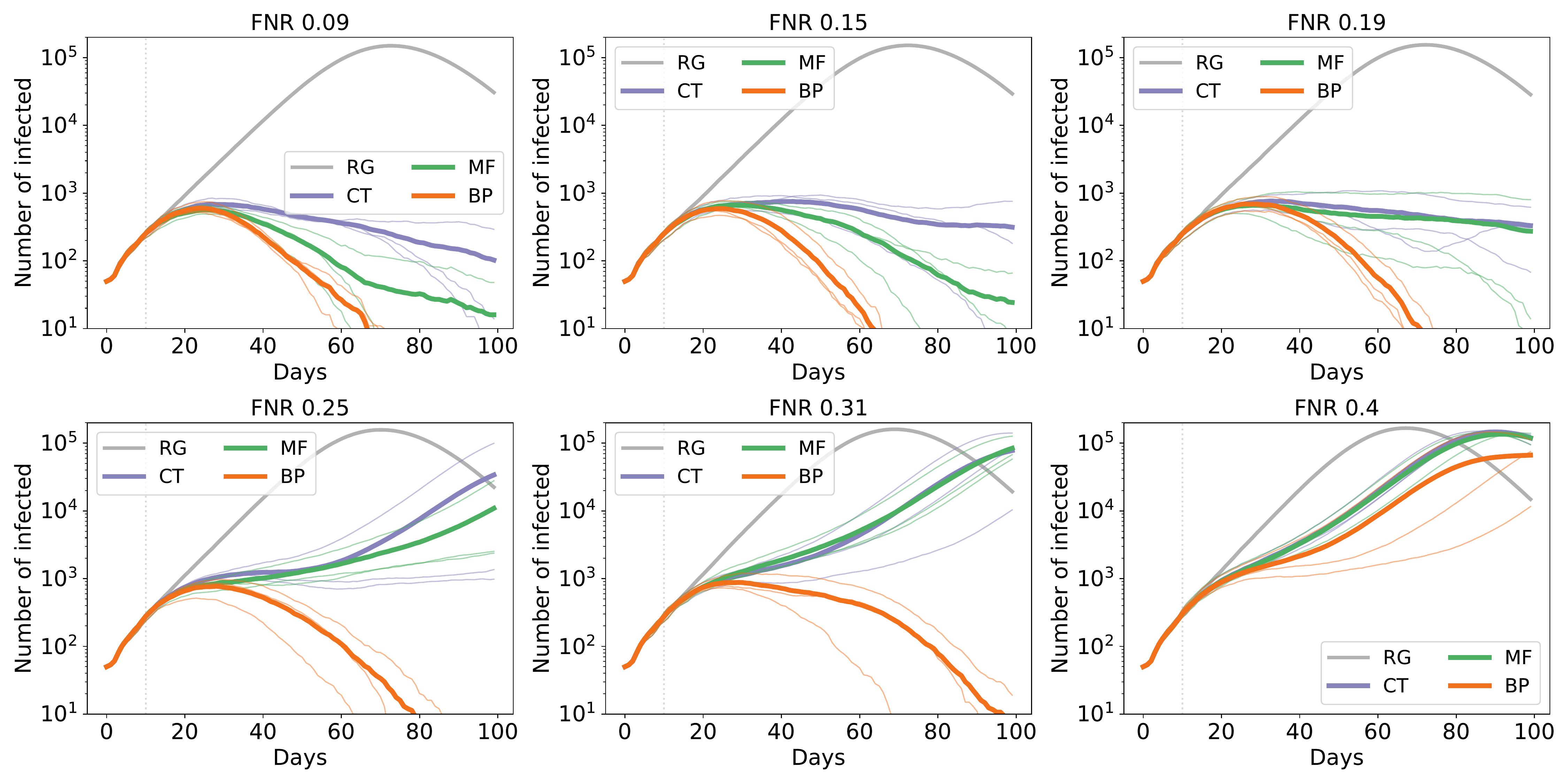}
    \caption{Effect of tests inaccuracy to the evolution of the controlled epidemics. We simulate the same intervention protocol as Figure \ref{fig:res_obs_1a} for 2500 daily observations (bottom panel). We consider here the effects of an additional source of noise, that is a non negligible false negative rate (FNR) of the results of the medical tests, from 0.09 to 0.40.\label{fig:res_3a_fnr}}
\end{figure}

We first consider the case in which the results of the medical tests are inaccurate and therefore there exists a fraction of the tested individuals incorrectly identified as uninfected or infected. Concerning the fraction of false positive tests this simply puts a small additional fraction of individual in isolation, but does not lead to deterioration of the epidemic control. We hence focus on the influence of false negatives and test how the performance depends on the false negative rate (FNR) of the medical tests. We remark that within the Bayesian framework it is possible to correctly include this information in a straight-forward way and we do so for the BP algorithm, but not for the MF as we want to keep it as simple as possible and test its robustness. 
In Figure \ref{fig:res_3a_fnr} we show the results of several simulations (three different realizations of the dynamics), starting from the setting in Figure \ref{fig:res_obs_1a} with 2500 medical tests and the quarantine of the entire households, when the FNR spans the range $[0.09, 0.40]$. We see that all the control strategies present good robustness with respect to the false negative tests. The CT controls the spreading up to FNR 0.19, MF is robustly better than CT up about FNR 0.25. Remarkably the intervention based of BP predictions not only limits the number of the infected individuals in time, but its performance remains almost unchanged with respect to the noiseless case as it completely stops the epidemic spreading even for large values of the FNR, up to 0.31. Figure \ref{fig:res_3b_fnr} of the Supplemental Material displays the results of the same experiment but starting from the initial setting of Figure \ref{fig:res_obs_1b} using 1000 medical tests, bottom panel. 
These results confirm the effectiveness of inference-based ranking and reveal the robustness in a realistic scenario of a fully Bayesian approach such as the proposed Belief Propagation in which the probabilistic evidence of tests can be consistently included as priors.
 
\begin{figure}
    \centering
    \includegraphics[width=1.0\linewidth]{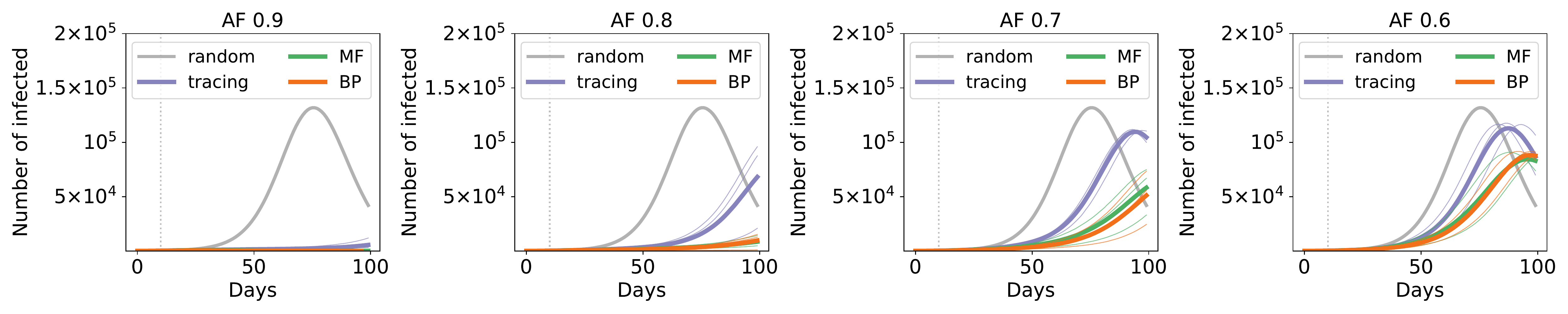}
    \caption{Effect of a poor adoption fraction of the mobile application to the number of infected individuals. We simulate the same intervention protocol as Figure \ref{fig:res_obs_1a} for 5000 daily observations. We assume here that only a fraction of the population, from 90 \% to 60\%, uses the mobile application for contact tracing.\label{fig:res_2a_app}}
\end{figure}

We now turn to the study of partial adoption of the mobile application. This is done in the simulation by hiding the contacts of a fraction of individuals (which are unknown to the inference algorithm): these hidden contacts represent individuals without the application or without smartphone. Figure \ref{fig:res_2a_app} shows the result of mitigation, in the OpenABM model, with AF (the fraction of individuals who have adopted the app) ranging between 0.6 and 0.9. It shows that the method is still effective in presence of partially detected contacts.
Although performance is severely affected, one observes that even at AF equal to 0.6 the use of inference algorithms allows to delay the spreading of the epidemic and to flatten the peak of infected individuals, way more efficiently than the classical contact tracing strategy. Furthermore, it should be noted that application utilization may be positively correlated to the number of contacts of individuals. Including more detailed information about mobile application utilization e.g. in population age classes may greatly reduce the impact of low adoption.
 Similar results are presented in the Supplemental Material, Figure \ref{fig:res_2b_app} for a smaller number of daily observations.

\paragraph{Conclusion}
\label{sec:conclusion}
The above results show that, in the regime where the epidemic is growing and exhaustive testing of all contacts is unfeasible, inference methods allow to contain the epidemics more efficiently than the classical tracing of contacts, which itself is better than random testing. Both inference schemes require exchange of information between individuals during a limited time window after they have been in contact, and could be implemented in contact-tracing smartphone applications in a distributed way. Additionally, numerical tests show that the approach is robust to false negatives in the test results as well as to partial adoption of the mobile tracing applications, although the adoption rate required for efficient control of the epidemic with the number of daily tests considered is large relative to the one of the currently deployed applications. 

The volume of daily exchanged messages per pair of individuals in the two proposed methods is constant with respect to both the population size and time (a total of about 1kB for MF, 1MB for BP per individual assuming $\sim10$ daily contacts). This volume is negligible when compared with normal data usage. A privacy-preserving implementation will require an additional overload, but the computational burden on the phone's CPU will remain negligible.

Having access to the estimated posterior probability of being infected in time, a series of threshold values could be put in place so as to suggest actions on individuals, including reduction of contacts, self-isolation and testing.  

With regard to privacy, it is worth emphasizing that the proposed inference methods are in principle more protective than the usual (manual) tracing. On the one hand, both can be implemented in a fully distributed way using point to point cryptography without fully centralized processing and storage of information on infections or contacts. On the other hand, by identifying individuals who have the largest probability of being infected through a cumulative process by which information is integrated, the direct attribution of potential infection events to a given individual is made much harder. Details of such fully privacy preserving implementation, along the lines of \cite{troncoso2020decentralized}, are left for future work. 

\paragraph{Acknowledgments} 

We would like to thank the ELLIS network for organizing a series of COVID-19 related workshops. We also want to thank Yoshua Bengio, Irina Rish, the MILA team working on the contact tracing problem, as well as Luca Ferretti and Ivan Bestvina for numerous enlightening discussions. We acknowledge computational resources provided by HPC@POLITO (\url{http://www.hpc.polito.it}) and well as Google Cloud for the SIPAR grant in COVID-19 research credits program. This work has been partially supported by the SmartData@PoliTO (\url{http://smartdata.polito.it}) center on Big Data and Data Science, the  French Agence Nationale de la Recherche under
grant ANR-17-CE23-0023-01 PAIL and from the Chaire CFM-ENS on data science. 

%
%

%
%
%
%
%
%

%
%
%
%
%
%
%
%
%
%
%
%
%
%
%
%
%
%
%
%
%
%
%
%
%

%
%
%
%
%
%

%
%
%
%
%
%

%
%

%

%
%
%
%

%
%

%
%
%
%
%
%
%
%
%
%
%
%
%
%
%
%
%
%
%

%

%

%

%

%
%
%
%
%
%
%
%

%
%
%
%
%

%
%
%
%
%

%
%
%
%
%

%
%
%
%
%

\bibliographystyle{unsrt}

\begin{thebibliography}{10}

\bibitem{Ferrettieabb6936}
Luca Ferretti, Chris Wymant, Michelle Kendall, Lele Zhao, Anel Nurtay, Lucie
  Abeler-D{\"o}rner, Michael Parker, David Bonsall, and Christophe Fraser.
\newblock Quantifying sars-cov-2 transmission suggests epidemic control with
  digital contact tracing.
\newblock {\em Science}, 368(6491), 2020.

\bibitem{bay2020bluetrace}
Jason Bay, Joel Kek, Alvin Tan, Chai~Sheng Hau, Lai Yongquan, Janice Tan, and
  Tang~Anh Quy.
\newblock Bluetrace: A privacy-preserving protocol for community-driven contact
  tracing across borders.
\newblock {\em Government Technology Agency-Singapore, Tech. Rep}, 2020.
\newblock
  \url{https://bluetrace.io/static/bluetrace\_whitepaper-938063656596c104632def383eb33b3c.pdf}.

\bibitem{AppleGoogle}
Privacy-preserving contact tracing.
\newblock https://www:apple:com/covid19/contacttracing/, 2020.

\bibitem{troncoso2020decentralized}
Carmela Troncoso, Mathias Payer, Jean-Pierre Hubaux, Marcel Salath{\'e}, James
  Larus, Edouard Bugnion, Wouter Lueks, Theresa Stadler, Apostolos Pyrgelis,
  Daniele Antonioli, et~al.
\newblock Decentralized privacy-preserving proximity tracing.
\newblock {\em arXiv preprint arXiv:2005.12273}, 2020.

\bibitem{alsdurf2020covi}
Hannah Alsdurf, Yoshua Bengio, Tristan Deleu, Prateek Gupta, Daphne Ippolito,
  Richard Janda, Max Jarvie, Tyler Kolody, Sekoul Krastev, Tegan Maharaj,
  et~al.
\newblock Covi white paper.
\newblock {\em arXiv preprint arXiv:2005.08502}, 2020.

\bibitem{chan2020pact}
Justin Chan, Shyam Gollakota, Eric Horvitz, Joseph Jaeger, Sham Kakade,
  Tadayoshi Kohno, John Langford, Jonathan Larson, Sudheesh Singanamalla, Jacob
  Sunshine, et~al.
\newblock Pact: Privacy sensitive protocols and mechanisms for mobile contact
  tracing.
\newblock {\em arXiv preprint arXiv:2004.03544}, 2020.

\bibitem{cho2020contact}
Hyunghoon Cho, Daphne Ippolito, and Yun~William Yu.
\newblock Contact tracing mobile apps for covid-19: Privacy considerations and
  related trade-offs.
\newblock {\em arXiv preprint arXiv:2003.11511}, 2020.

\bibitem{raskar2020apps}
Ramesh Raskar, Isabel Schunemann, Rachel Barbar, Kristen Vilcans, Jim Gray,
  Praneeth Vepakomma, Suraj Kapa, Andrea Nuzzo, Rajiv Gupta, Alex Berke, et~al.
\newblock Apps gone rogue: Maintaining personal privacy in an epidemic.
\newblock {\em arXiv preprint arXiv:2003.08567}, 2020.

\bibitem{yedidia2003understanding}
Jonathan~S Yedidia, William~T Freeman, and Yair Weiss.
\newblock Understanding belief propagation and its generalizations.
\newblock {\em Exploring artificial intelligence in the new millennium},
  8:236--239, 2003.

\bibitem{Lokhov2014}
Andrey~Y. Lokhov, Marc Mézard, Hiroki Ohta, and Lenka Zdeborová.
\newblock Inferring the origin of an epidemic with a dynamic message-passing
  algorithm.
\newblock {\em Physical Review E}, 90(1), 07 2014.

\bibitem{lokhov2015dynamic}
Andrey~Y Lokhov, Marc M{\'e}zard, and Lenka Zdeborov{\'a}.
\newblock Dynamic message-passing equations for models with unidirectional
  dynamics.
\newblock {\em Physical Review E}, 91(1):012811, 2015.

\bibitem{altarelli_bayesian_2014}
Fabrizio Altarelli, Alfredo Braunstein, Luca Dall{\textquoteright}Asta,
  Alejandro Lage-Castellanos, and Riccardo Zecchina.
\newblock Bayesian inference of epidemics on networks via belief propagation.
\newblock {\em Physical Review Letters}, 112(11):118701, March 2014.

\bibitem{altarelli_patient-zero_2014}
Fabrizio Altarelli, Alfredo Braunstein, Luca Dall{\textquoteright}Asta,
  Alessandro Ingrosso, and Riccardo Zecchina.
\newblock The patient-zero problem with noisy observations.
\newblock {\em Journal of Statistical Mechanics: Theory and Experiment},
  2014(10):P10016, October 2014.

\bibitem{altarelli_containing_2014}
F.~Altarelli, A.~Braunstein, L.~Dall{\textquoteright}Asta, J.~R. Wakeling, and
  R.~Zecchina.
\newblock Containing epidemic outbreaks by message-passing techniques.
\newblock {\em Physical Review X}, 4(2):021024, May 2014.

\bibitem{braunstein_inference_2016}
Alfredo Braunstein and Alessandro Ingrosso.
\newblock Inference of causality in epidemics on temporal contact networks.
\newblock {\em Sci Rep}, 6:27538, June 2016.

\bibitem{bindi2017predicting}
Jacopo Bindi, Alfredo Braunstein, and Luca Dall’Asta.
\newblock Predicting epidemic evolution on contact networks from partial
  observations.
\newblock {\em Plos one}, 12(4):e0176376, 2017.

\bibitem{braunstein_network_2019}
Alfredo Braunstein, Alessandro Ingrosso, and Anna~Paola Muntoni.
\newblock Network reconstruction from infection cascades.
\newblock {\em Journal of The Royal Society Interface}, 16(151):20180844,
  February 2019.

\bibitem{Ellis_conf}
https://ellis.eu/covid-19/events/ellis-against-covid-19-06-05-2020, 5 2020.

\bibitem{ViraTrace}
\url{https://github.com/ViraTrace/InfectionModel}, 4 2020.

\bibitem{herbrich2020crisp}
Ralf Herbrich, Rajeev Rastogi, and Roland Vollgraf.
\newblock Crisp: A probabilistic model for individual-level covid-19 infection
  risk estimation based on contact data.
\newblock {\em arXiv preprint arXiv:2006.04942}, 2020.

\bibitem{hinch_2020}
Hinch R, Probert W, Nurtay A, Kendall M, Wymant C, Hall M, Lythgoe K, Cruz~A B,
  Zhao L, Stewart A, Ferretti L, Bonsall~D Abeler-Dorner~L, and Fraser C.
\newblock Covid-19 agent-based model with instantaneous contract tracing.
\newblock Technical report, 2020.

\bibitem{epidemicmitigation}
https://github.com/sibyl-team/epidemic\_mitigation, 2020.

\bibitem{2020BentoutParameterEstimation}
Soufiane Bentout, Abdennasser Chekroun, and Toshikazu Kuniya.
\newblock Parameter estimation and prediction for coronavirus disease outbreak
  2019 (covid-19) in algeria.
\newblock 7:306--318, 05 2020.

\bibitem{franco2020covid19}
Nicolas Franco.
\newblock Covid-19 belgium: Extended seir-qd model with nursery homes and
  long-term scenarios-based forecasts from school opening, 2020.

\bibitem{fintzi2020using}
Jonathan Fintzi, Damon Bayer, Isaac Goldstein, Keith Lumbard, Emily Ricotta,
  Sarah Warner, Lindsay~M. Busch, Jeffrey~R. Strich, Daniel~S. Chertow,
  Daniel~M. Parker, Bernadette Boden-Albala, Alissa Dratch, Richard Chhuon,
  Nichole Quick, Matthew Zahn, and Vladimir~N. Minin.
\newblock Using multiple data streams to estimate and forecast sars-cov-2
  transmission dynamics, with application to the virus spread in orange county,
  california, 2020.

\bibitem{KEFAYATI2020}
Sarah KEFAYATI, Hu~Huang, Prithwish Chakraborty, Fred Roberts, Vishrawas
  Gopalakrishnan, Raman Srinivasan, Sayali Pethe, Piyush Madan, Ajay Deshpande,
  Xuan Liu, Jianying Hu, and Gretchen Jackson.
\newblock On machine learning-based short-term adjustment of epidemiological
  projections of covid-19 in us.
\newblock {\em medRxiv}, 2020.

\bibitem{lorch_spatiotemporal_2020}
Lars Lorch, William Trouleau, Stratis Tsirtsis, Aron Szanto, Bernhard
  Schölkopf, and Manuel Gomez-Rodriguez.
\newblock A {Spatiotemporal} {Epidemic} {Model} to {Quantify} the {Effects} of
  {Contact} {Tracing}, {Testing}, and {Containment}.
\newblock {\em arXiv:2004.07641 [physics, q-bio, stat]}, 2020.

\bibitem{ferretti_quantifying_2020}
Luca Ferretti, Chris Wymant, Michelle Kendall, Lele Zhao, Anel Nurtay, Lucie
  Abeler-Dörner, Michael Parker, David Bonsall, and Christophe Fraser.
\newblock Quantifying {SARS}-{CoV}-2 transmission suggests epidemic control
  with digital contact tracing.
\newblock {\em Science}, 368(6491), May 2020.

\bibitem{mezard2009information}
Marc Mezard and Andrea Montanari.
\newblock {\em Information, physics, and computation}.
\newblock Oxford University Press, 2009.

\bibitem{pearl1982reverend}
Judea Pearl.
\newblock Reverend bayes on inference engines: a distributed hierarchical
  approach.
\newblock In {\em Proceedings of the Second AAAI Conference on Artificial
  Intelligence}, pages 133--136, 1982.

\bibitem{altarelli_large_2013}
Fabrizio Altarelli, Alfredo Braunstein, Luca Dall'Asta, and Riccardo Zecchina.
\newblock Large deviations of cascade processes on graphs.
\newblock {\em Physical Review E}, 87(6):062115, June 2013.

\bibitem{altarelli_optimizing_2013}
F.~Altarelli, A.~Braunstein, L.~Dall{\textquoteright}Asta, and R.~Zecchina.
\newblock Optimizing spread dynamics on graphs by message passing.
\newblock {\em Journal of Statistical Mechanics: Theory and Experiment},
  2013(09):P09011, September 2013.

\end{thebibliography}

\newpage
\appendix
\section{Brief description of the OpenABM model}\label{app:OpenABM}
The OpenABM computational model by R. Hinch et al.~\cite{hinch_2020} involves
simulating epidemic propagation in an age-stratified population based
on the UK national census data (see Table 1 in \cite{hinch_2020}). Each
individual is represented by a node of a multi-layered network and
takes part in three different subnets describing different social
contexts: two static subnets represent households and workplaces while
a degree-heterogeneous random network, different every day, represents
the occasional interactions which individuals have on a daily basis.
Age stratification influences both the composition of households (the
elderly tend to live with other elderly, children preferably live
with young adults) and the social activity level of individuals (e.g.
participation in the workplace network, average number of random daily
interactions). The epidemic states considered by the model are those
discussed above, apart from the exposed state, which is not explicitly
modeled. Hospitalization and/or death is possible only for severely
symptomatic patients, while the resistant state is eventually reached
in case of no or mild symptoms. The epidemic dynamics is modeled as
a discrete-time stochastic process, with a temporal resolution of
one day, in which infected individuals can transmit the disease with
a (daily) infection rate which depends mainly on the symptomatic state
of the potential infector, the age of the potential infected and the
time passed since the potential infector became infected. The latter
dependence, modeled through a Gamma distribution, is an attempt to
describe the temporal variation of the infectiousness level of SARS-CoV-2.
In particular, the incubation period (usually represented by the E
state), is here implicitly taken into account by assuming an infectiousness
level which does not immediately grow from zero. Finally, the infection
rate depends on an intrinsic infectiousness level of the virus and on
the type of network on which the contact occurs. Although the notion
of duration of contacts is not present, the infectiousness rate associated
with contacts inside households is larger than for the other environments,
to account for the typically longer duration of domestic interactions.
With the exception of the viral transmission process just described,
all other possible transitions between individual epidemic states
(e.g. transition from mild symptomatic infected state to the resistant
state) are independent of the state of the other individuals. These
events are characterized by a (discrete) waiting time, also distributed
according to a Gamma distribution or, in case of dichotomous support,
according to a shifted Bernoulli distribution. The parameters of these
distributions have been extracted from recent SARS-CoV-2 literature
and are summarized in Tables 5-6-7 of Hinch et al. \cite{hinch_2020}.

The model provides the possibility of intervention in order to slow
down and, if possible, contain the epidemic outbreak. In particular,
it is possible to introduce interventions of increasing severity,
from case-based measures (e.g. quarantine for individuals which are
positive to swab tests and their housemates) to mobility restrictions
for some categories of individuals and lockdown scenarios. In this
respect, the OpenABM model is very appropriate for the implementation of
contact tracing strategies. Finally, the OpenABM model also provides for the possibility of varying the adoption fraction of the contact tracing app within the population, possibly introducing different percentages of adoption in different age groups of individuals.

\section{Mean-Field Inference of Risk}\label{app:MFdetails}

\label{sec:MFapproach}
The mean field inference  is based on a prior epidemic model which is an agent-based SIR model. 
At each time, an individual $i$ is in either one of the three states: {Susceptible, Infected, Removed}, $x_i(t) \in\{S,I,R\}$ (where the "Removed" state means either dead, or recovered and having acquired immunity).

 When going from time $t$ to time $t+1$, the following events can take place:
\begin{itemize}
    \item If $x_i(t)=R$: $x_i(t+1)=R$. \\
    \item If $x_i(t)=I$:  we call $\mu_i$ the probability that an infected individual $i$ recovers: \\$ \begin{cases}x_i(t+1)&=I \quad \text{with probability }1-\mu_i\\
    x_i(t+1)&=R \quad \text{with probability }\mu_i
    \end{cases}$
    \item If $x_i(t)=S$: One looks at all the individuals $j$ which are infected and have been in contact with $i$ at time $t$ (in practice this means between day $t$ and day $t+1$). Define this set of individuals as $\partial i(t)$. Each individual $j$ in $\partial i(t)$ infects $i$ with a probability $\lambda_{j \to i}(t)$. More precisely:\\
    $ \begin{cases}x_i(t+1)&=I \quad \text{with probability }1-\prod_{j\in \partial i(t)}\left(1-\lambda_{j\to i}(t)\right)\\
    x_i(t+1)&=S \quad \text{with probability }\prod_{j\in \partial i(t)}\left(1-\lambda_{j\to i}(t)\right)
    \end{cases}$
\end{itemize}
To resume, the parameters of the model are:
\begin{itemize}
    \item $\mu_i$: The probability of removal of the infectious patient $i$. \\
    \item $\lambda_{i \to j}(t)$: The transmission probability given that there was a contact between an infected $i$ and susceptible $j$ at time $t$. This depends on $i,j$ and on $t$, depending on the duration and nature of contacts between $i$ and $j$ between day $t$ and $t+1$ transmission probability changes with contact time, as well as the sanitary measures, such as wearing masks. 
\label{eq:params}
\end{itemize}

From these rules one obtains the propagation of the epidemic $p\left(x_{i}^{t}|x_{\partial i}^{t-1},x_{i}^{t-1}\right)$ as defined in eq.~(\ref{eq:markov}).

In order to understand and monitor the propagation of the epidemic, and to develop mitigating strategies, it is important to evaluate the probabilities that individual $j$ is in state $S$, $I$ or $R$ at a given time $t$. We denote these probabilities respectively by $P_S^j (t)$, $P_I^j(t)$, $P_R^j (t)$. 

These \emph{marginal} probabilities are in general difficult to evaluate. A straightforward strategy for this evaluation is to simulate a large number of instances of the propagation, and estimate $P_S^j (t)$ as the fraction of instances where individual $j$ is in state $S$ at time $t$. However this requires a centralized system, and a large computing power. Here, instead, we use some approximate techniques from statistical physics, that allow for a good estimate of $P_S^j (t)$ through a fully distributed method, using only simple exchange of information at each contact.

The \emph{Mean-Field} (MF) method  computes the marginal probabilities of~(\ref{eq:markov}) through an iterative process.
The probability of individual $j$ receiving the infection from her contact $k$ at time $t$ depends on their contact transmission  $\lambda_{k \to j}(t)$ and on the joint probability of $j$ being $S$ and $k$ being $I$ at time $t$. 
The mean field approximation estimates this joint probability by the product $P_{S}^{j}(t) P_I^k(t) $.
Using this approximation, one can write the probability that individual $j$ is susceptible at time $t+1$ as
\begin{align}
P_{S}^{j}(t+1)&=P_{S}^{j}(t)\prod_{k\in\partial j(t)}\left(1-P_I^k(t)\lambda_{k \to j}(t)\right).
\end{align}
The probability of being recovered is
\begin{align}
P_{R}^{j}(t+1)&=P_{R}^{j}(t)+\mu_{j}P_{I}^{j}(t)
\end{align}
and the probability of being infected is obtained using the fact that $P_S+P_R+P_I=1$.
In practice, considering that the probability of transmission is small, our MF algorithm is based on the following linearized form for $P_S$ (and we have checked that this linearization is fine in the regimes of epidemic propagation that we explore):
\begin{align}
P_{S}^{j}(t+1)&=P_{S}^{j}(t)\left(1-\sum_{k\in \partial j(t)}P_I^k(t)\lambda_{k \to j}(t)\right)\notag,\\
P_{R}^{j}(t+1)&=P_{R}^{j}(t)+\mu_{j}P_{I}^{j}(t),\notag\\
P_{I}^{j}(t+1)&=P_{I}^{j}(t) + P_{S}^{j}(t)\sum_{k\in \partial j(t)}P_I^k(t)\lambda_{k \to j}(t) - \mu_j P_{I}^{j}(t).
\label{eq:MFequations}
\end{align}

The  mean-field equations have an intuitive content which is easy to understand. They basically reproduce in an agent-dependent model the equations used for the global monitoring of the proportions of S,I,R states in a population. They can also be derived as a limiting case of the dynamical message passing equations from \cite{Lokhov2014}, when the transmission probabilities are small.

This approach offers several advantages. Every individual $j$ can estimate her probabilities $P_S^j (t)$, $P_I^j(t)$, $P_R^j (t)$ every day, by updating the equations (\ref{eq:MFequations}). These probabilities can be stored in her phone. For the update, individual $j$ needs to receive, during the contact with $k$, the information on $\lambda_{k \to j}(t)$ and the information from $k$ about his estimates of $P_S^k (t)$, $P_I^k(t)$, $P_R^k (t)$. The value of $\lambda_{k \to j}(t)$ is the standard contact-tracing information, which estimates the encounter duration within a certain distance, as used in all contact tracing applications that are being developed, for instance based on bluetooth signals between the phones of $j$ and $k$. On top of this, the phone of individual $k$ should send the values of $P_S^k (t)$ and  $P_I^k(t)$ to individual $j$ during the contact, and reciprocally. The information is fully distributed, there is no need for a central system that stores the full information, and the data exchange can be encrypted.
We suppose that, at time $t_{obs}$ an individual $i$ is tested or presents illness-associated syndromes. 
Then the state of $i$ is known: $x_i^{obs}(t)\in\{S,I,R \}$ and $P_q^i(t_{obs})=\delta_{q,q^{obs}_i(t)}$. In case of syndromes at time $t_{obs}$ the probability $P_q^i(t_{obs})$ is updated
on the basis of external medical data, namely the probability to be infected among all people presenting the same set of syndromes.

A simple inference method that turns out to be quite efficient consists in adapting the mean-field equations (\ref{eq:MFequations}) in order to take into account the results of tests and symptoms. The information about tests and syndroms must be propagated back in time and be used to update the risk levels of the contacts of person $i$ in recent times. Assume that we are estimating the probabilities for each individual $i$ to be in each of the three states $q$ at a given time $t$, $P_q^i(t)$. We run the mean-field equations (\ref{eq:MFequations}) starting at time $t-t_{MF}$ with the whole population $S$, and imposing the constraints due to the tests done in the interval $[t-t_{MF},t]$ as follows. If $j$ is tested at a time $t_{obs}$ in this interval, then:
\begin{align}
    &\text{if } x_j(t_\text{obs}) = S:\quad
    P_S^j(t') = 1 \quad \text{for } t' \in [t-t_{MF}, t_\text{obs}]  \\
    &\text{if } x_j(t_\text{obs}) = I:\quad
    P_I^j(t') = 1 \quad \text{for } t_\text{obs} - \tau \leq t' \leq t_\text{obs}  \\
    &\text{if } x_j(t_\text{obs}) = R:\quad
    P_R^j(t') = 1 \quad \text{for } t' \geq t_\text{obs} 
\end{align}
Our inference procedure depends on two parameters:  $\tau$ is the  typical time between the infection and the testing consecutive to the apparition of syndroms, and $t_{MF}$ is the integration time of the mean-field procedure.

%

%
%
%

%
%
%
%
%
%
%

%
%
%
%
%
%
%
%
%
%

%
%
%
%
%
%
%

%
%
%
%
%
%
%
%
%
%
%
%
%
%
%
%

%
%
%
%

%

%
%
%
%
%
%

%
%
%
%

%

%

%
%
%
%
%
%
%
%
%
%
%

%
%
%
%
%
%
%
%

%
%

%
%
%
%
%
%
%
%

%
%

%

%

%
%
%
%
%
%
%
%

%

%
%
%
%
%
%
%
%
%
%
%
%
%
%
%
%
%
%

%

%
%
%
%

\section{Belief Propagation Inference of Risk}\label{app:BPdetails}
\subsection{Graphical model setting}

Each pair $i,j$ of individuals will be in mutual contact in a finite
set of instants $X_{ij}\subset\mathbb{R_{\infty}}=\mathbb{R}\cup\left\{ +\infty\right\} $.
For reasons that will become clear in the following, we will always
assume $\infty\in X_{ij}$. As time advances, instantaneous contagion
will happen with probability $\lambda$ at time $s_{ij}\in X_{ij}$
if $i$ is infected and $j$ is susceptible. We will assume $\lambda=\lambda_{ij,}^{s_{ij},t_{i}}$
to possibly depend both on the specific (absolute) contact time $s_{ij}$,
on the direction of the contact and on the time $t_{i}$ of infection
of individual $i$. Individual $i$ can thus become infected in one
instant in the set $X_{i}=\cup_{j\in\partial i}X_{ij}$. We will denote
by $t_{i}\in X_{i},r_{i}\in\mathbb{R}$ respectively the times of
infection and recovery of individual $i$, with $t_{i}=\infty$ (resp.
$r_{i}=\infty$) if the individual did not become infected (resp.
recovered) within the time-frame. 

We will assume the recovery delay $r_{i}-t_{i}$ of node $i$ to be
distributed with a continuous distribution with pdf $p_{R,i}\left(r_{i}-t_{i}\right)$.
We will assume a set of factorized, site-dependent observations $p_{O,i}\left(\mathcal{O}_{i}|t_{i},r_{i}\right)$.
Model parameters will be hidden for the moment inside functions $p_{R,i}$,
$\lambda_{ij}^{s,t}$, and $p_{O,i}$ and we will include them only
later to avoid cluttering the notation.

The standard SIR model can be obtained by setting $p_{R,i}\left(r_{i}-t_{i}\right)=\mu e^{-\mu\left(r_{i}-t_{i}\right)}$,
$\lambda_{ij}^{s,t}\equiv\lambda$. In this setup, the model is memory-less (Markov) on the state of infection variables $x_{i}^{t}\in\left\{ S,I,R\right\} $.

In the following we will always assume that $t_{i}\in X_{i}$ and
$s_{ij},s_{ji}\in X_{ij}$. Given the times of infection and recovery
$t_{i}$ and $r_{i}$, the transmission delay $s_{ij}$ has ``truncated''
generalized geometric distribution 
\begin{eqnarray}
S_{ij}\left(s_{ij}|t_{i},r_{i}\right) & =\mathbb{I}\left[t_{i}<s_{ij}<r_{i}\right] & \lambda_{ij}^{s_{ij},t_{i}}\prod_{t_{i}<s<s_{ij}}\left(1-\lambda_{ij}^{s,t_{i}}\right)+\mathbb{I}\left[s_{ij}=\infty\right]\prod_{s\geq r_{i}}\left(1-\lambda_{ij}^{s,t_{i}}\right)
\end{eqnarray}
because $i$ will be infectious in the open time interval $\left(t_{i},r_{i}\right)$
and will possibly (i.e. if $j$ is susceptible at the time) transmit
the disease at some time $s_{ij}$ in that interval if he did not
transmit it before. If time $r_{i}$ arrives and he did not transmit
the disease, the individual will recover and never transmit the disease
through that link; then transmission time on that link will be nominally
$s_{ij}=\infty$. Special attention must be taken for auto-infections
(otherwise, no infection can enter into the closed system). By adding
a contact with an additional extra always-infected virtual neighbor
$i^{t*}$ to node $i$ at time instant $t\in X_{i}$ with probability
$p_{t}\left(s_{i^{t*}i}=t\right)=\gamma_{i}^{t}$ (typically small),
$p_{t}\left(s_{i^{t*}i}=\infty\right)=1-\gamma_{i}^{t}$, $i$ will
spontaneously self-infect at time $t$ with probability $\gamma_{i}^{t}$
(if it is susceptible at that time\footnote{A particularly interesting case is with $\gamma_{i}^{0}=\gamma\to0$
and $\gamma_{i}^{t}=0$ for $t>0$: in this case individuals can be
self-infected only at time $0$, representing a closed system with
a single unknown seed at time $t=0$.}). Let us define~$A_{i}\left(\boldsymbol{s}_{i^{*}}\right)=A_{i}\left(\left\{ s_{i^{t*}i}\right\} _{t\in X_{i}}\right)=\prod_{t\in X_{i}\setminus\left\{ \infty\right\} }p_{t}\left(s_{i^{t*}i}\right)$.
For convenience, let us define $\partial^{*}i$ the enlarged neighborhood
of $i$ including all extra nodes $\left\{ i^{t*}\right\} _{t\in X_{i}}.$
Given $\left\{ s_{ki}\right\} $ for $k\in\partial^{*}i$, the infection
time $t_{i}$ satisfies, in a deterministic way:

\begin{eqnarray}
t_{i} & = & \min_{k\in\partial^{*}i}s_{ki},
\end{eqnarray}

We can now write the joint pdf of discrete variables $\boldsymbol{t},\boldsymbol{s}$
and continuous variables~$\boldsymbol{r}$ as
\begin{eqnarray}
p\left(\boldsymbol{t},\boldsymbol{r},\boldsymbol{s}\right) & \propto & \prod_{i}\delta\big(t_{i},\min_{k\in\partial^{*}i}s_{ki}\big)A_{i}\left(\boldsymbol{s}_{i^{*}}\right)R_{i}\left(r_{i}-t_{i}\right)\prod_{\left(ij\right)}S_{ij}\left(s_{ij}|t_{i},r_{i}\right)
\end{eqnarray}
so then as $p\left(\boldsymbol{t},\boldsymbol{r},\boldsymbol{s}|\mathcal{O}\right)\propto p\left(\mathcal{O}|\boldsymbol{t},\boldsymbol{r}\right)p\left(\boldsymbol{t},\boldsymbol{r},\boldsymbol{s}\right)$,
we get
\begin{eqnarray}
p\left(\boldsymbol{t},\boldsymbol{r},\boldsymbol{s}|\mathcal{O}\right) & \propto & \prod_{i}\delta\big(t_{i},\min_{k\in\partial^{*}i}s_{ki}\big)A_{i}\left(\boldsymbol{s}_{i^{*}}\right)R_{i}\left(r_{i}-t_{i}\right)p_{O,i}\left(\mathcal{O}_{i}|t_{i},r_{i}\right)\prod_{\left(ij\right)}S_{ij}\left(s_{ij}|t_{i},r_{i}\right)\label{eq:boltzman}
\end{eqnarray}

A note about $r_{i}$s: all terms in \eqref{eq:boltzman} except $R_{i}$
are constant as functions of $r_{i}$ in any interval $\left(\hat{r}_{i},\hat{r}_{i}'\right)$
of consecutive times in $X_{i}$. We will exploit this fact and write
$r_{i}=\hat{r}_{i}+u_{i}$, where $\hat{r}_{i}=\max\left\{ r\in X_{i}:r<r_{i}\right\} $.
Integrating away the $u_{i}$ variables in \eqref{eq:boltzman}, we
obtain an all-discrete model for variables $\boldsymbol{t},\boldsymbol{s},\boldsymbol{\hat{r}},$
with an identical expression to \eqref{eq:boltzman} but in which
$r_{i}$ has been replaced by $\hat{r}_{i}$ and $R_{i}\left(r_{i}-t_{i}\right)$
by $\hat{R}_{i}\left(\hat{r}_{i}-t_{i}\right)=\int_{\hat{r}_{i}-t_{i}}^{\hat{r}_{i}'-t_{i}}p_{R,i}\left(u\right)du.$
For simplicity of notation, we will drop the $\hat{}$ symbols in
the following.

\subsection{Belief propagation equations}

A naive interpretation of \eqref{eq:boltzman} as a graphical model
would introduces many unneeded short cycles that were not present
in the original contact network. For example, pairs $\left(t_{i},s_{ji}\right),\left(t_{i},s_{ij}\right),\left(t_{j},s_{ij}\right),\left(t_{j},s_{ji}\right)$
share respectively factors with indices $i,\left(ij\right),j,\left(ji\right)$
, effectively forming a small cycle. A simple solution consists in
regrouping factors as in \eqref{eq:fg} and considering $\left(s_{ij},s_{ji}\right)$
as a single variable:
\begin{eqnarray}
\Psi_{i}\left(t_{i},r_{i},\left\{ s_{ki},s_{ik}\right\} _{k\in\partial i}\right) & = & \delta\big(t_{i},\min_{k\in\partial^{*}i}s_{ki}\big)A_{i}\left(\boldsymbol{s}_{i^{*}}\right)R_{i}\left(r_{i}-t_{i}\right)\prod_{j\in\partial i}S_{ij}\left(s_{ij}|t_{i},r_{i}\right)\label{eq:fg}
\end{eqnarray}
which results in a factor graph for \eqref{eq:boltzman} in which
variables $\left(s_{ij},s_{ji}\right)$ have degree two and live in
the middle of the original edges, and vars $t_{i},r_{i}$ have degree
$1$, i.e. a topology that closely follows the one of the original
contact network:
\begin{eqnarray}
p\left(\boldsymbol{t},\boldsymbol{r},\boldsymbol{s}|\mathcal{O}\right) & = & \frac{1}{Z}\prod_{i}\Psi_{i}\label{eq:gmodel}
\end{eqnarray}

The corresponding BP equations for $\Psi_{i}$ are

\begin{eqnarray}
m_{ij}\left(s_{ij},s_{ji}\right) & \propto & \sum_{t_{i}}\sum_{r_{i}}p_{O,i}\left(\mathcal{O}_{i}|t_{i},r_{i}\right)A_{i}\left(\boldsymbol{s}_{i^{*}}\right)R_{i}\left(r_{i}-t_{i}\right)S_{ij}\left(s_{ij}|t_{i},r_{i}\right)\times\label{eq:mij}\\
 &  & \times\sum_{\left\{ s_{ki}\right\} }\delta\big(t_{i},\min_{k\in\partial^{*}i}s_{ki}\big)\prod_{k\in\partial^{*}i\setminus j}S_{ik}\left(s_{ik}|t_{i},r_{i}\right)m_{ki}\left(s_{ki},s_{ik}\right)\nonumber 
\end{eqnarray}
 and marginals for $t_{i}$ are
\begin{eqnarray}
b_{i}\left(t_{i}\right) & \propto & \sum_{r_{i}}p_{O,i}\left(\mathcal{O}_{i}|t_{i},r_{i}\right)A_{i}\left(\boldsymbol{s}_{i^{*}}\right)R_{i}\left(r_{i}-t_{i}\right)S_{ij}\left(s_{ij}|t_{i},r_{i}\right)\times\label{eq:marginal}\\
 &  & \times\sum_{\left\{ s_{ki}\right\} }\delta\big(t_{i},\min_{k\in\partial^{*}i}s_{ki}\big)\prod_{k\in\partial^{*}i}S_{ik}\left(s_{ik}|t_{i},r_{i}\right)m_{ki}\left(s_{ki},s_{ik}\right)
\end{eqnarray}
and similarly for $r_{i}.$ A more efficient computation of the equations
can be achieved by defining:
\begin{align*}
G_{k}^{0}\left(t_{i},r_{i}\right) & =\sum_{\substack{s_{ki}\geq t_{i}\\
s_{ik}>t_{i}
}
}S_{ik}\left(s_{ik}|t_{i},r_{i}\right)m_{ki}\left(s_{ki},s_{ik}\right)\\
G_{k}^{1}\left(t_{i},r_{i}\right) & =\sum_{\substack{s_{ki}>t_{i}\\
s_{ik}>t_{i}
}
}S_{ik}\left(s_{ik}|t_{i},r_{i}\right)m_{ki}\left(s_{ki},s_{ik}\right)
\end{align*}
and substituting the extra neighbor message
\[
m_{i^{*t}i}\left(s_{i^{*t}i},s_{ii^{*t}}\right)=\begin{cases}
\gamma_{i}^{t} & s_{i^{*t}i}=t,s_{ii^{*t}}=\infty\\
1-\gamma_{i}^{t} & s_{i^{*t}i}=\infty,s_{ii^{*t}}=\infty
\end{cases}
\]
Equations \eqref{eq:mij}, \eqref{eq:marginal} can be then rewritten
as
\begin{align}
m_{ij}\left(s_{ij},s_{ji}\right)\propto & \sum_{t_{i}}\sum_{r_{i}}p_{O,i}\left(\mathcal{O}_{i}|t_{i},r_{i}\right)A_{i}\left(\boldsymbol{s}_{i^{*}}\right)R_{i}\left(r_{i}-t_{i}\right)S_{ij}\left(s_{ij}|t_{i},r_{i}\right)\times\label{eq:bpeff}\\
 & \times\sum_{\left\{ s_{ki}\right\} }\left(\prod_{j\in\partial^{*}i}\mathbb{I}\left[s_{ki}\geq t_{i}\right]-\prod_{j\in\partial^{*}i}\mathbb{I}\left[s_{ki}>t_{i}\right]\right)\prod_{k\in\partial^{*}i\setminus j}S_{ik}\left(s_{ik}|t_{i},r_{i}\right)m_{ki}\left(s_{ki},s_{ik}\right)\nonumber \\
\propto & \sum_{t_{i}<s_{ji}}\sum_{r_{i}\geq t_{i}}p_{O,i}\left(\mathcal{O}_{i}|t_{i},r_{i}\right)R_{i}\left(r_{i}-t_{i}\right)S_{ij}\left(s_{ij}|t_{i},r_{i}\right)\times\\
 &\times\bigg\{\prod_{k\in\partial i\setminus j}G_{k}^{0}\left(t_{i},r_{i}\right)-\left(1-\gamma_{i}^{t_{i}}\right)\prod_{k\in\partial i\setminus j}G_{k}^{1}\left(t_{i},r_{i}\right)\bigg\}\nonumber + \\
 & +\sum_{r_{i}\geq s_{ji}}p_{O,i}\left(\mathcal{O}_{i}|s_{ji},r_{i}\right)R_{i}\left(r_{i}-s_{ji}\right)\prod_{k\in\partial i\setminus j}G_{k}^{0}\left(s_{ji},r_{i}\right)\nonumber \\
b_{i}\left(t_{i}\right)\propto & \sum_{r_{i}}p_{R,i}\left(r_{i}-t_{i}\right)p_{O,i}\left(\mathcal{O}_{i}|t_{i},r_{i}\right)\times\\
 & \bigg\{\prod_{k\in\partial i}G_{k}^{0}\left(t_{i},r_{i}\right)-\left(1-\gamma_{i}^{t_{i}}\right)\prod_{k\in\partial i}G_{k}^{1}\left(t_{i},r_{i}\right)\bigg\}\nonumber \\
b_{i}\left(r_{i}\right)\propto & \sum_{t_{i}}p_{R,i}\left(r_{i}-t_{i}\right)p_{O,i}\left(\mathcal{O}_{i}|t_{i},r_{i}\right)\times\\
 & \bigg\{\prod_{k\in\partial i}G_{k}^{0}\left(t_{i},r_{i}\right)-\left(1-\gamma_{i}^{t_{i}}\right)\prod_{k\in\partial i}G_{k}^{1}\left(t_{i},r_{i}\right)\bigg\}\nonumber 
\end{align}

Note that products $\prod_{k\in\partial i\setminus j}G_{k}$ can be
computed simultaneously for $j\in\partial i$ in time $O\left(\left|\partial i\right|\right)$
(either by computing it as the fraction $G_{j}^{-1}\prod_{k\in\partial i}G_{k}$,
or by first recursively computing $\prod_{\ell=1}^{\ell'}G_{k_{\ell}}$
and $\prod_{\ell=\ell'}^{\left|\partial i\right|}G_{k_{\ell}}$ for
$\ell'=1,\dots,\left|\partial i\right|$ and then $\prod_{\ell\neq\ell'}G_{k_{\ell}}=\prod_{\ell=1}^{\ell'-1}G_{k_{\ell}}$$\prod_{\ell=\ell'+1}^{\left|\partial i\right|}G_{k_{\ell}}$,
a method that does not involve divisions and is thus numerically more
stable). The resulting implementation of the update of all messages
in factor $\Psi_{i}$ has complexity $O\left(\left|X_{i}\right|\sum_{j\in\partial i}\left(\left|X_{i}\right|+\left|X_{ij}\right|^{2}\right)\right)$
.

\subsection{Finite-window approximation}

In the BP-based epidemic tracing scheme, exchanged messages between two individuals grow quadratically with the number of temporal contacts occurred between them. However, only recent contacts are important to determine marginal probabilities at present time, therefore keeping only a short time window (about two or three weeks) is sufficient to obtain quasi-optimal results. For better accuracy, information about contacts and observations at the dropped times is included approximately as simple factorized priors applied at the start of the window. This prior contains the posterior probability at the first non-dropped time computed only using contacts and observations at the dropped time (and the prior computed in the previous step). All simulations have been performed using a $21$ days time window. 

\subsection{Algorithm Parameters}\label{subsec:BPparams}

For the OpenABM model, we chose to use Gamma distributions for the recovery density $p_{R.i}=p_R$ and a rescaled Gamma for the infection transmissivity $\lambda_{ij}^{s_{ij},t_i}=p_I(s_{ij}-t_i)$. The five parameters were fitted from experimental data produced by the model (parameters could in principle be also learned or adjusted online during the process through an approximate maximum likelihood procedure \cite{altarelli_patient-zero_2014}, but we leave this for future work). Note that the model used for inference with BP is still much simplified with respect to OpenABM itself, in particular having only three states (against 11 in OpenABM). As a consequence, results are only weakly sensitive to the parameters. The used values were Gamma($k$ = 10, $\mu = 0.57$) for $p_R$ and Gamma($k$=5.76, $\mu=0.96$), with a scale of 0.25 (multiplied by 2 for intra-household contacts as with MF inference) for $\lambda$.  
The self-infection probability was chosen to be $p_{seed}=1/N$ at time $t=0$ ($k/N$ where $k$ is the number of patient zeros would  bring slightly better results, but would use inaccessible information) and 0 for $t>0$ except for the case with partial adoption, in which we allowed a small probability of self-infections at times $t>0$ to avoid plain incompatibility between the inference model and observations due to undetected transmissions). The parameter $\delta_{rank}$ for the computation of the ranking was chosen to be 10 days.

\label{app:BPparameters}


\section{Additional Results}
\label{app:AdditionalResults}

In this section, we present additional results to stress how the containment measures associated with the inference-based methods proposed in the main text, are effective in limiting the epidemics. We consider a realistic spreading dynamics given by the OpenABM model \cite{hinch_2020} and, unlike the setting illustrated in the main text, we study the case in which the restrictions are applied earlier in time (i.e. after a week from the beginning of the epidemics) and the size of the epidemics at the initial time is smaller (the number of patients zero here is 20), being also consistent with an earlier intervention scenario. 

\begin{figure}[tb]
\centering
		\includegraphics[width=0.8\linewidth]{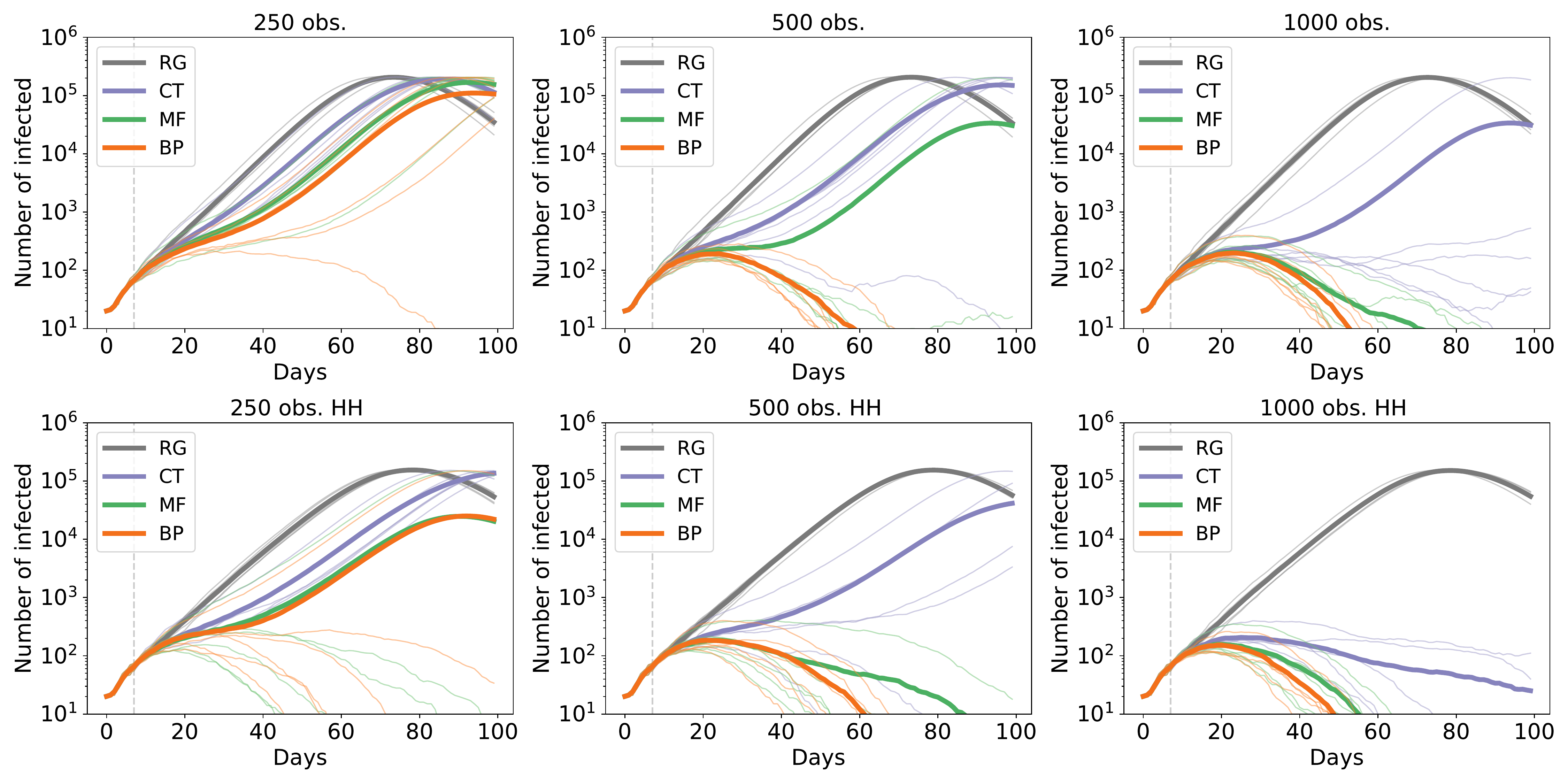}
	\caption{
	Effect of the control strategy on the epidemic spreading. In all the panels we show the number of infected individuals in a time window of 100 days when some intervention is applied starting from day 7. The number of patients zero here is set to 20. Thin lines represent the results for single instances of the epidemics, while the think line is the average among the different realizations. We compare the effect of an increasing number of available medical tests per day (from left to right), performed to the individuals at risk suggested by the app. The top panels depict a scenario where only tested positive individuals are confined, limiting their contacts to the cohabitants, while the bottom panels show how the number of infected individuals change if the entire household is quarantined whenever an infected is detected. 
	\label{fig:res_obs_1b} 
	}
\end{figure}

In Figure \ref{fig:res_obs_1b}, we display the behavior of the number of infected individuals as a function of time, similarly to Figure \ref{fig:res_obs_1a} of the main text. 
Qualitatively, we retrieve the same behavior we have observed in Figure \ref{fig:res_obs_1a}: inference-based ranking allows for a more effective intervention resulting in a remarkable decrease of the number of infected individuals, and when the number of tests is sufficiently large, the epidemics are stopped in slightly more than two months. Quantitatively, we notice that to control the spreading a reduced number of daily medical tests are needed (about ten times less than those used for the results in Figure \ref{fig:res_obs_1a}), suggesting that early intervention is equally effective with a more parsimonious usage of testing resources. 

\begin{figure}[tb]
\centering
		\includegraphics[width=0.8\linewidth]{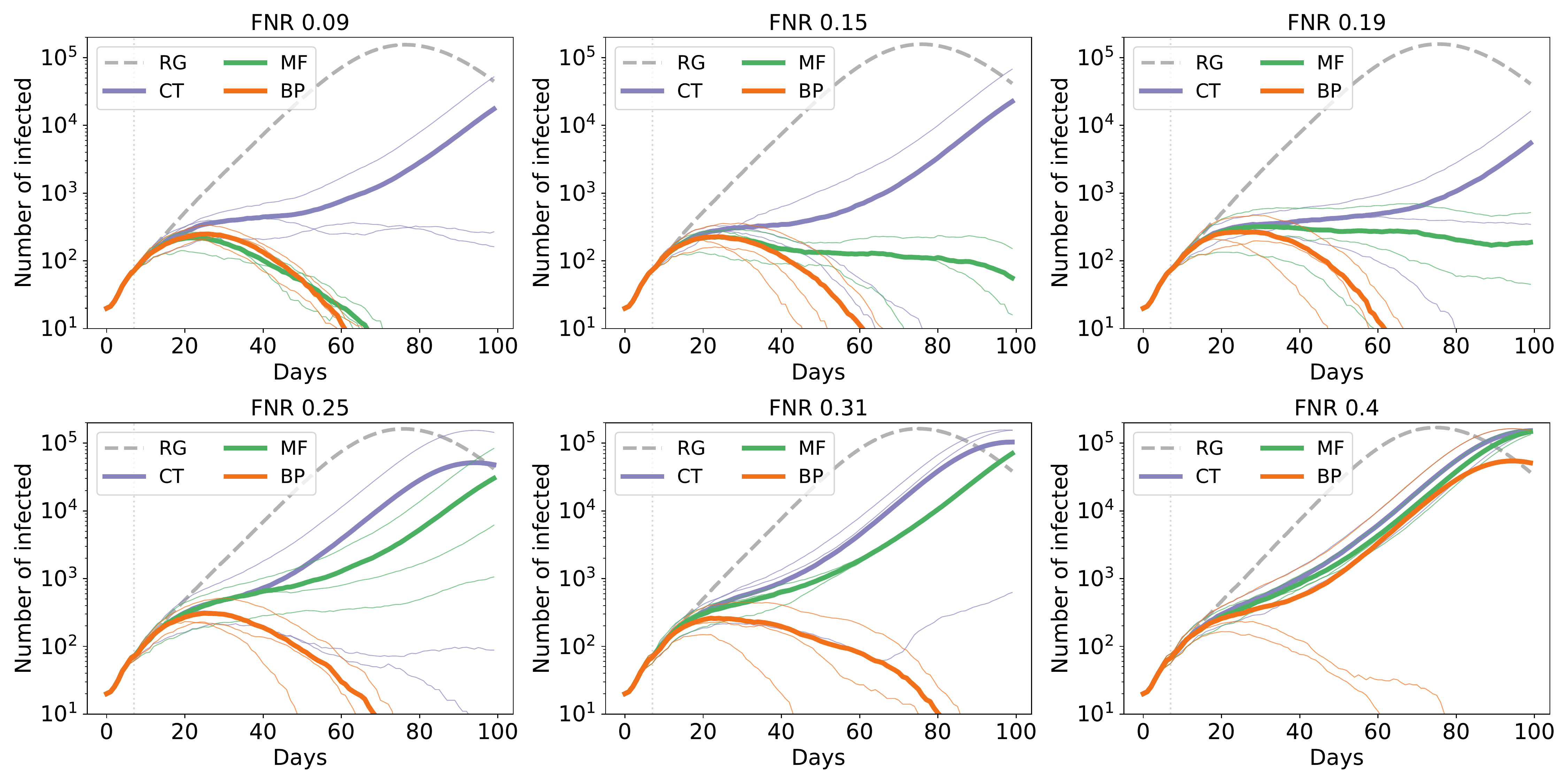}
	\caption{Effect of tests inaccuracy to the evolution of the controlled epidemics. We simulate the same intervention protocol as Figure \ref{fig:res_obs_1b} for 1000 daily observations (bottom panel). We consider here the effects of an additional source of noise, that is a non negligible false negative rate (FNR) of the results of the medical tests, from 0.09 to 0.40.\label{fig:res_3b_fnr}}
\end{figure}

\begin{figure}[bt]
   \centering
    \includegraphics[width=1.0\linewidth]{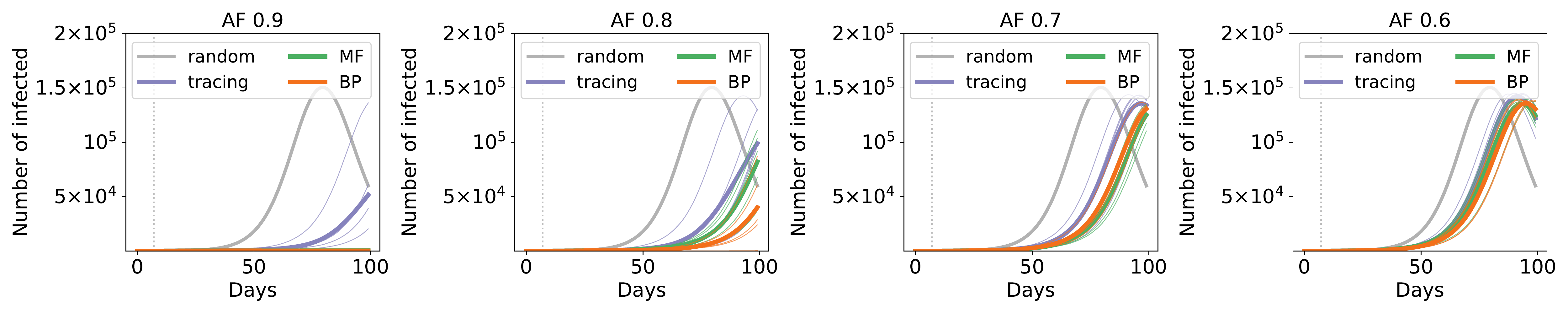}
   \caption{Effect of a poor adoption fraction of the app to the number of infected individuals. We simulate the same intervention protocol as Figure \ref{fig:res_obs_1b} for 1000 daily observations. We assume here that only a fraction of the population, from 90 \% to 60\%, uses the app for contact tracing.\label{fig:res_2b_app}
   }
\end{figure}

Figure \ref{fig:res_3b_fnr} suggests how robust the containment measures are when the medical tests are inaccurate. We quantify their sensitivity through a false-negative rate (FNR) of the medical tests (similar results are discussed in the main text, see Figure \ref{fig:res_3a_fnr}). The setting is the same as in Figure \ref{fig:res_obs_1b} when 1000 medical tests are performed to the individuals who present the higher risk to be infected, according to RG, CT, MF, and BP. These curves support the conclusion of the main text: inference-based methods successfully contain an epidemic in a realistic setting with non-zero FNR. Indeed, CT and MF are able to contain the epidemics up to FNR equal to 0.09 and 0.25, respectively. BP shows the most robust performance as it is able to stop the epidemics even for larger values of the FNR, up to 0.31.

Finally, we show the average epidemic size as a function of the time in Figure \ref{fig:res_2b_app} when only a fraction of the population uses the app for contact tracing and, therefore, only partial information on potentially infectious contacts is available to the risk assessment methods. Up to an adoption fraction of 0.7, inference-based methods (MF and BP) allows for a significantly slowing down of the epidemic spreading compared to both the random and CT-based testing.


\end{document}